# Slope of Dry Granular Materials Surface is Generally Curved

Fathan Akbar[1], Elfi Yuliza[2], Nadya Amalia[3], Handika Dany Rahmayanti[4],

and Mikrajuddin Abdullah[1,a]

[1]Department of Physics, Bandung Institute of Technology,
Jalan Ganesa 10, Bandung 40132, Indonesia
[2]Department of Physics, Faculty of Mathematics and Natural Sciences, University of Bengkulu,
Jl. WR. Supratman, Bengkulu 38371, Indonesia
[3]Research Center of Materials Science and Metallurgy, National Research and Innovation Agency, South Tangerang 15314, Indonesia
[4]State Polytechnic of Creative Media, Jalan Srengseng Sawah, Jakarta 12630, Indonesia

[a]Corresponding author: mikrajuddin@gmail.com

**Abstract**

As if it has become a consensus that the slope of the granular heap is straight, and from that, an angle of repose is defined. However, closer inspection shows that the slopes are not truly straight, instead, more often they have convex shapes, although with small convexity. Similar curvature is observed on the sand dunes' surfaces. We derive equations together with experiments on 10 types of granular materials to show that, in general, the slope of the granular heap is convex while the straight slope is a critical state that rarely occurs. The angle of repose is allowed to vary within a narrow range around a critical value. If the angle of repose is slightly greater (smaller) than the critical value, the slope is convex (concave). The model also answers why the surfaces of granular heaps, including sand dunes, are more often convex rather than concave or straight, and advocate the simultaneous estimation of the coefficient of friction on the grain and the type of packing in the heap. The results might open new understandings related to granular material properties as well as sand transport and geological records in several deserts.

Exotic images of sand dunes like that in the Namib Sand Sea [1,2,3,4,5,6] where some of the world's largest linear dunes are situated [7], Erg Chebbi and Erg Chigaga [8], Great Sand Dune National Park [9,10], and Badain Jaran Desert [11,12] generally show the curvature surface (**Supplementary 1**). The sand dunes are gigantic sand heaps with the angle of repose ranges between 30° and 35° [13,14], similar to that of dry sands [15]. The sand dunes are examples of aeolian deposits which are unique geomorphological and paleoclimatic proxies [16,17]. Its shape has remained a long-standing open issue, of relevance for several areas of aeolian research and planetary sciences [18], and also facilitate the recognition in geological records [7].

Similarly, the surfaces of the granular heaps are generally not straight (**Supplementary 2**). Figure 4 in [19] clearly shows a slightly convex profile around the base of the silver sand heap. By simulation using 50000 particles with various aspect ratios, Zhao et al [20] produced a slightly convex surface of the granular heap. The maximum height of grain piles as a function of distance from the vertical line that passes through the peak shows quite scattered data [21], but they can be fit with a convex curve. Figure 1 in [22] clearly shows the convex surface of the heap of snow (the snow is often treated as a granular material [23,24]). Images of pellet heaps reported in [25] also show a convex profile.

The formation of grain heaps is of paramount importance to understand complex cooperative phenomena [26]. The heaps are of great interest for bulk solids characterization [27]. Variations in the heap's angle of repose were investigated as a function of experimental parameters and deviations in the shape of the tail of sandpiles were discovered [28]. By numerical investigation of very large granular heap, Topić et al [26] showed the change in angle of repose against the radius from the center, the finding that can only be explained by considering that the slope of heap is curved. However, there are no reports on claiming that the curving of the granular heap slope is an existing state. In this paper, we demonstrate that the slope of the granular heap can have three distinct profiles: convex, concave, and straight. Experiments on ten types of granular materials were conducted to test the model.

**Figure 1** on the left is the illustration of the arrangement of particles on a heap having a height of $H$. The $x$ (horizontal) axis is selected in the direction of the height and the $y$ (vertical) axis is parallel to the baseline. These directions were chosen for convenience. We define the angle of repose as the angle at the upper corner made by a straight line from the baseline to the top. The angle of this line to the horizontal axis, $\bar{\theta}$. We hypothesize the heap slope deviates from this straight line as can be observed in sand dunes [29] or granular materials heap (for examples, Figs. 1 and 2 in [23], Fig. 7 in [30]).

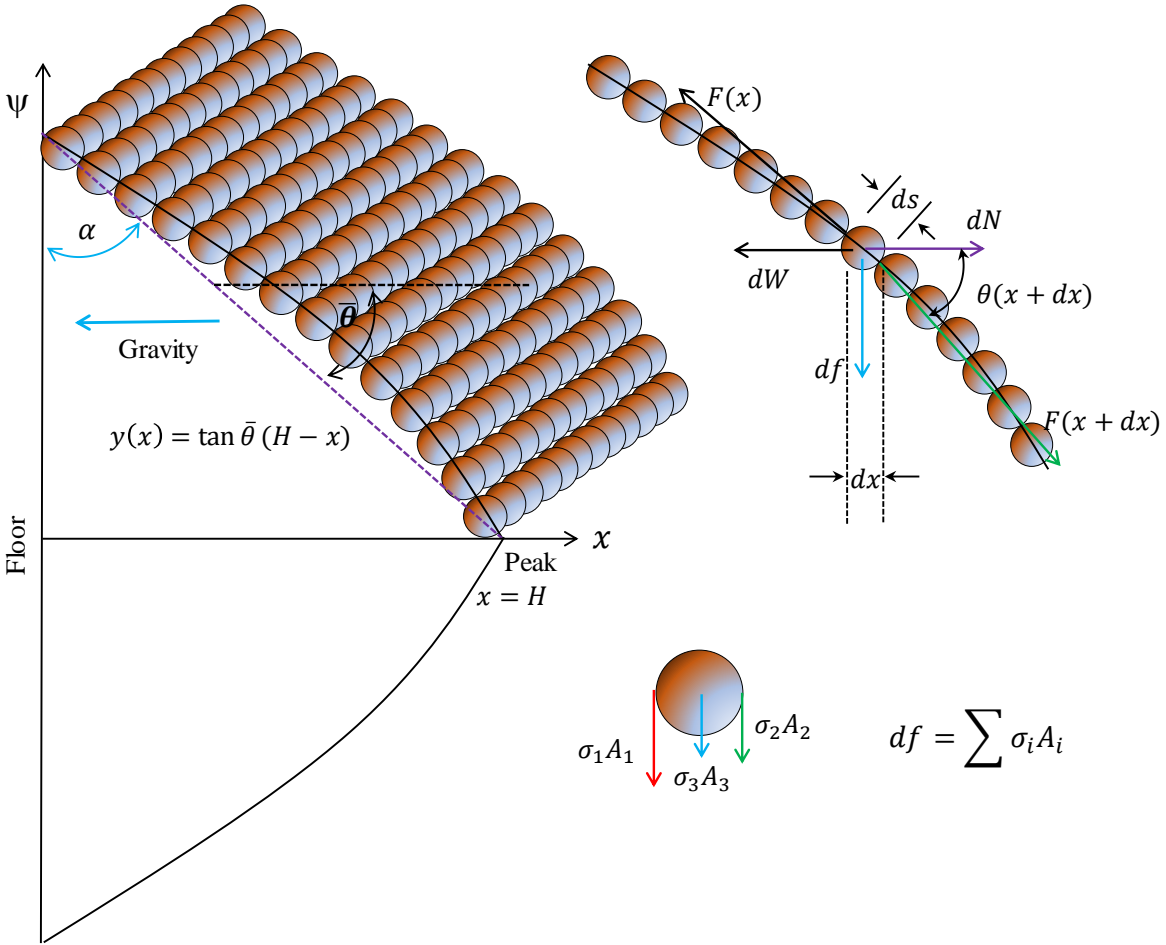

**Figure 1** Illustration of a granular heap surface. The coordinates $(H, 0)$ are the top of the heap and $(0, H/\tan \alpha)$ as the position of the bottom edge where is the average static angle of repose. The average slope of the heap surface is $\tan \bar{\theta}$ where $\bar{\theta} - \alpha = -\pi/2$.

We focus only on the arrangement of the particles in the top surface layer by assuming that the arrangement on this layer determines the granular surface profile. Similar considerations have been used by other authors [31,32]. Jaeger et al [33] described when the slope angle exceeds the angle of repose, only the top layers of sand flow freely downhill. According to the image during the avalanche, when the slope angle exceeds the angle of repose, only a few layers of particles on the surface flow [34]. We focus on a single layer with the thickness in the z-direction is equal to the particle diameter so that the system is similar to a curved string [35].

Consider one element at a position between $x$ to $x + dx$ which contains only one particle. The horizontal and vertical forces equilibrium are $dN = -d/dx[F(x)\cos\theta(x)]dx + dW$ and $df = -d/dx[F(x)\sin\theta(x)]dx$, respectively. The frictional force $df$ comes from stresses on the various contact sides of the particle and we assume that it is directed only inward, the magnitude of which is $|df| = |\sum \sigma_i A_i|$ where $\sigma_i$ is the stress on the i-th contact surface and $A_i$ is the i-th contact surface area. When the frictional force is maximum, the stress is assumed proportional to the normal force, $\sigma_i = \kappa_i dN$, so that $|df| = |\sum \kappa_i A_i dN| = \mu|dN|$, where $\mu = \sum \kappa_i A_i$ is the coefficient of friction. Referring to the Coulomb equation [36], this approach was taken with the assumption that the cohesive forces are neglectable [37]. From forces equilibrium equations one gets

$$\left| -\frac{d}{dx}[F(x)\sin\theta(x)]dx \right| = \left| -\mu \frac{d}{dx}[F(x)\cos\theta(x)]dx + \mu dW \right| \quad (1)$$

The coefficient of friction generally comes from the sliding and rolling of the particles. We have assumed that only the sliding motion contributes mainly based on the fact that the rolling torque associated with rolling friction is proportional to particle radius [38] so that for very small particle sizes its contribution can be neglected.

The force $F(x)$ acting along the surface slope can be approximated with a hydrostatic-like force [39,40,41] and all other forces are accommodated in the frictional force. The hydrostatic-like pressure satisfies $P(x) = \rho g(H - x)$ where $\rho$ is the density of the granular system after taking into account the presence of pores and $g$ is the gravity acceleration. The hydrostatic force at $x$ is $F(x) = \rho g A(H - x)$ where $A = \pi R^2$ is the effective surface area.

The heap slope equation reads $\psi(x) = y(x) + z(x)$ where $y(x) = (H - x)\tan\bar{\theta}$ is the equation of the straight slope and $z(x)$ is the deviation from it, $|z(x)| \ll |y(x)|$ and $|dz/dx| \ll |dy/dx|$. By several mathematical steps (see **Suppelemntary 3**) we get deviation of the slope with respect to the line $y(x)$ satisfies

$$\frac{z(x)}{H} = \frac{b}{1+a}\left(\left(1-\frac{x}{H}\right)^{-a} - \left(1-\frac{x}{H}\right)\right). \tag{2}$$

where

$$a = -\frac{\mu\gamma\cos\alpha}{\rho g A(-\mu\cos\alpha+\sin\alpha)} \tag{3}$$

$$b = \frac{\mu\sin\alpha+\cos\alpha}{-\mu\cos\alpha+\sin\alpha} - \frac{\mu\gamma}{\rho g A\sin\alpha(-\mu\cos\alpha+\sin\alpha)} \tag{4}$$

To ensure convergence of the solution, one needs $a < 0$, resulting in $-\mu\cos\alpha + \sin\alpha > 0$ (Eq. (3)) or $\mu < \tan\alpha$. The largest deviation appears at $x/H = 1 - (-a)^{1/(1+a)}$, that is

$$\frac{z_{max}}{H} = b(-a)^{-a/(1+a)}. \tag{5}$$

The deviations of 10 granular materials observed at present work (sesame, white sugar, coriander, rice, sand, lentils, mung bean, barley, dal bean, and basmati) are shown in **Figs. 2(a)** and **(b)** (see **Supplementary 4** for more details). We examined the images of heaps having the heights of around 10 cm. The particle sizes of all materials conform to a lognormal distribution (see **Supplementary 5**). The average particle size of each is: sesame ($L = 0.2761$ cm, $W = 0.1737$ cm), white sugar (0.0181 cm), coriander ($D = 0.3096$ cm), rice ($L = 0.516$ cm, $W = 0.226$ cm), beach sand ($D = 0.0012$ cm), lentil ($L = 0.475$ cm, $W = 0.275$ cm), mung bean ($L = 0.596$ cm, $W = 0.418$ cm), barley ($L = 0.492$ cm, $W = 0.336$ cm), dal bean ($L = 0.616$ cm, $W = 0.326$ cm), and basmati ($L = 0.845$ cm, $W = 0.186$ cm), where $L$ stands for length, $W$ for width, and $D$ for diameter. Symbols are the measurement results and the fitting curves have been obtained using Eq. (2). Parameters $a$ and $b$ have been estimated based on the position of the peak deviation and the peak value, so they are not free parameters. The measured and calculated data are comparable. The coefficients of determination ($R^2$) as shown in **Table 1** are also quite close to unity, indicating the fitting equation is adequate.

**Figures 2(c)** and **(d)** are comparisons of curves obtained from Eq. (2) after adding the values for the straight slope, $y(x)/H = \tan(\pi/2 - \alpha_0)(1 - x/H)$, to $z(x)/H$: (c) for the rice and (d) for the sesame. The curves obtained are very consistent with the observed data.

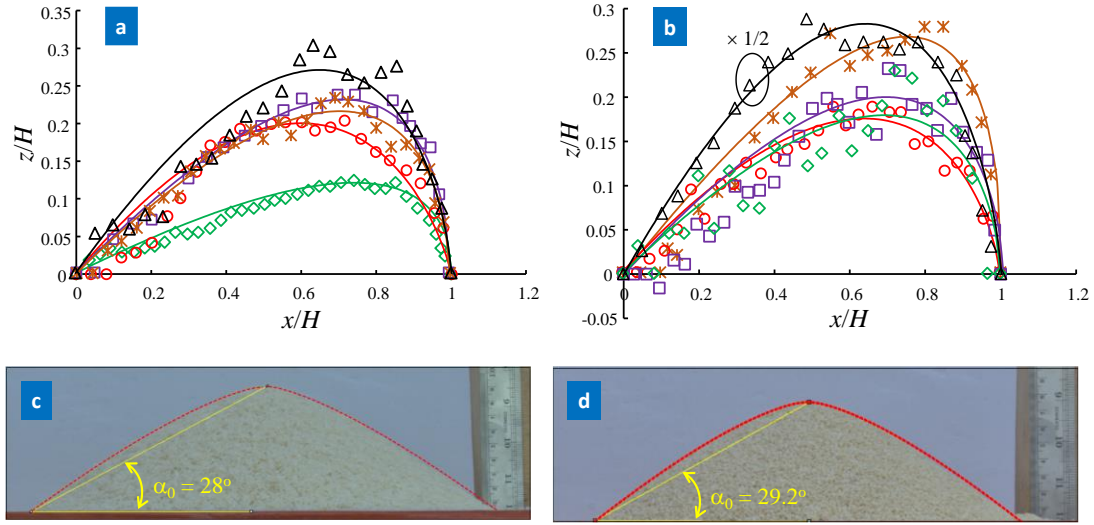

**Figure 2** (a) and (b) Comparison of calculating results (curves) and measurement results (symbols that have the same color) of the deviation of 10 granular materials. In (a): (triangles) coriander, (squares) sesame, (circle) white sugar, (stars) rice, and (diamonds) beach sand. In (b): (triangles, height has been multiplied by 1/2) lentil, (diamonds) mung bean, (circles) barley, (squares) dal bean, and (stars) basmati. (c) The image of rice heap with the calculated curve (sum of the corresponding curve in Fig. (a) and the heights belong to the straight surface) and (d) is the image of sesame heap and its corresponding fitting curve.

**Table 1** Measured data and fitting results for 10 types of granular materials.

| Granular materials | Measured repose angles, $\alpha_0$ (deg) | $a$ (fitting results) | $b$ (fitting results) | Positions of maximum deviation ($x/H$) | Maximum deviations ($z_{max}/H$) | $R^2$ |
|---|---|---|---|---|---|---|
| White sugar | 31.7 | -1.188 | 0.594 | 0.60 | 0.20 | 0.883 |
| Sesame | 29.2 | -0.623 | 0.503 | 0.715 | 0.23 | 0.979 |
| Sand | 33.9 | -0.498 | 0.240 | 0.75 | 0.120 | 0.917 |
| Coriander | 20.7 | -0.910 | 0.701 | 0.65 | 0.27 | 0.884 |
| Rice | 28.0 | -0.662 | 0.482 | 0.705 | 0.215 | 0.972 |
| Basmati | 25.5 | -0.5005 | 0.540 | 0.75 | 0.27 | 0.944 |
| Barley | 21.5 | -0.9544 | 0.4648 | 0.64 | 0.175 | 0.935 |
| Dal bean | 26.0 | -0.723 | 0.466 | 0.69 | 0.20 | 0.823 |
| Mung bean | 23.0 | -0.714 | 0.417 | 0.69 | 0.18 | 0.796 |
| Lentil | 16.0 | -0.957 | 1.516 | 0.64 | 0.57 | 0.959 |

*Estimation of friction coefficient.* If $\phi$ is the porosity of the material, one has $\rho = (1 - \phi)\rho_g$ where $\rho_g$ is the density of the granular material [42]. Referring to **Fig. 1**, we have $dW = \gamma ds = \gamma(2R)$. But $dW = (4/3)\pi R^3 \rho_g g$ so that $\gamma = (2/3)\pi R^2 \rho_g g$ and $\gamma/\rho g A = 2/3(1 - \phi)$. Based on Eqs. (4) and (5) one gets

$$\mu = \frac{\cos\alpha \sin\alpha}{\frac{2}{3(1-\phi)} - \frac{2}{3(1-\phi)}\left(\frac{b}{a}\right)\cos\alpha\sin\alpha - \sin\alpha\sin\alpha} \tag{6}$$

Equation (6) allows the estimation of the friction coefficient from the measurement of the angle of repose together with the fitting parameters $a$ and $b$. The square and diamond symbols in **Fig. 3(a)** were calculated using Eq. (6) and parameters in **Table 1**. The square symbols were calculated using the random closed packing (RCP, $\phi = 0.36$) and diamond symbols using the random loose packing (RLP, $\phi = 0.40$). It appears that the measured data are in agreement with the simulation results [38,43].

From **Table 1** we find the $b/a$ ratio varies from $-0.5$ to $-1.0$ after removing the data for lentils that differ considerably. We draw the curve $\alpha_0 - \mu$ at of $b/a = -0.5$ and $b/a = -1.0$ (**Fig. 3(a)**), where the symbols represent data on 10 granular materials and curves are the results of calculations with the Eq. (6) for RCP (dashed) and RLP (solid). The measurement results are mostly above the RCP curve, concluding that the surfaces of the 10 granular materials were convex to also indicate that packings inside all granular materials are RCPs. Therefore, the present model is able to estimate simultaneously the coefficient of static friction in the dry granular materials and the type of packing.

As shown in **Fig. 3(a)**, when we move from the dashed curve ($\phi = 0.36$) to the solid curve ($\phi = 0.40$) which has the same color, at a certain angle of repose, the coefficient of friction increases. An increase in $\phi$ means a decrease in the packing fraction, so we might state that with a decrease in packing, the friction coefficient increases. This prediction is in accordance with previous reports, either experimental [44], modeling [45], and simulation [46,47,48] that random loose packing decreases with increasing the friction coefficient.

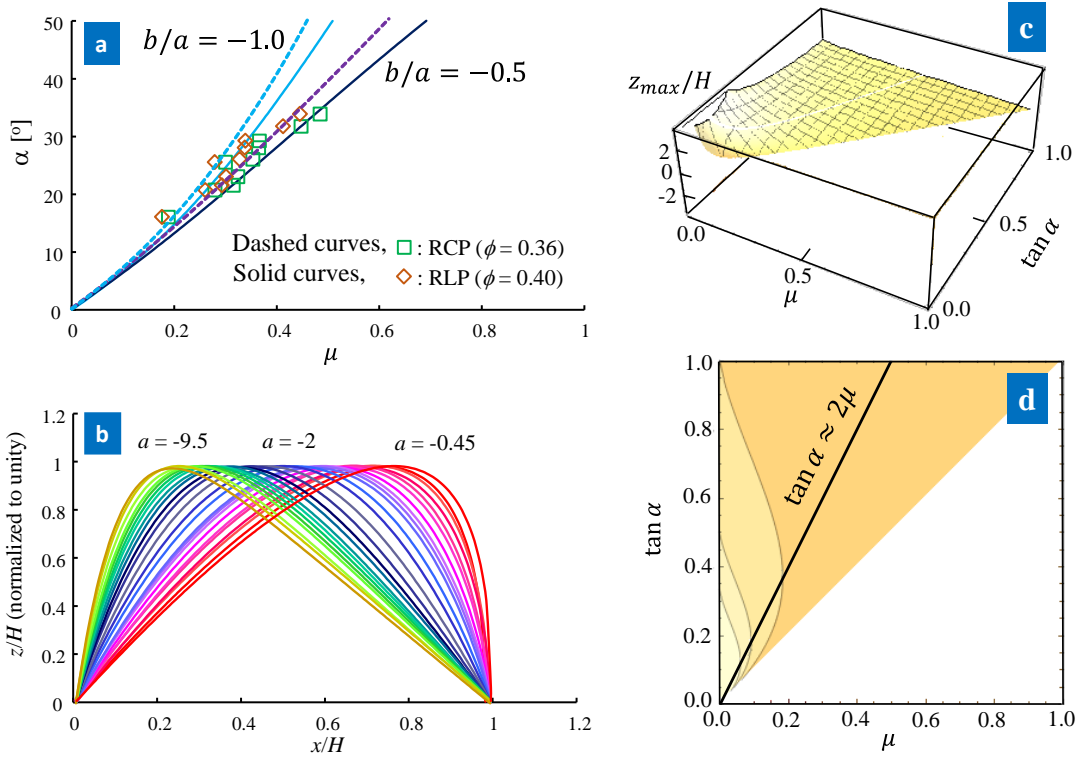

**Figure 3** (a) Angle of repose as a function of the coefficient of friction. The symbols are the results of calculations using parameters $a$ and $b$ in **Table 1**. These values can be considered experimental values. Calculations were carried out on all granular materials listed in **Table 1**. Calculations were performed for two packing fractions, RCP (squares) and RLP (diamonds). The curves are the results of calculations using Eq. (6). Based on the data in Table 1, we find that the $b/a$ ratio varies from -0.5 to -1.0 by excluding the ratio which is quite skewed for lentils. We perform calculations on these two limit values for RCP (dashed curves) and RLP (solid curves). (b) The shapes of the deviation curve at various parameter values $a$. The maximum height of the curve has been normalized to unity. We just want to show how the location of the peaks depends on parameter $a$. (c) and (d) Plot of $z_{max}/H$ at various values of $\mu$ and $\tan \alpha$ ((c): surface plot and (d): contour plot) in the area that satisfies $\tan \alpha > \mu$.

Equation (2) converges at $0 \leq x \leq H$ if $a < 0$, and based on Eq. (3) it requires $-\mu \cos \alpha + \sin \alpha > 0$ or $\mu < \tan \alpha$. It is easily shown that the divergence does not occur at $a = -1$ by writing $1 + a = \Delta a$ where $a \to -1$ and $|\Delta a| \to 0$, $z_{max}/H = b(1 - \Delta a)^{a/\Delta a} \approx b(1 - \Delta a)^{1/\Delta a} = b \exp(-\Delta a) \approx b$.

From Eq. (2), $\lim_{a \to 0^-} z_{max}/H = \lim_{a \to -\infty} z_{max}/H \to 0$. Referring to Eq. (3), $a \to 0^-$ only occurs if $\mu \to 0$ and $a \to -\infty$ only occurs in very critical conditions, namely $\mu = \tan \alpha$. However, we have shown $\mu < \tan \alpha$ so the condition $a \to -\infty$ cannot be achieved. Thus, to produce the straight slope, either $\mu \to 0$ or $b \to 0$ is mandatory. The condition that $\mu \to 0$ is trivial and need not be discussed. We only consider the condition of $b \to 0$. Since $a < 0$, $(-a)^{-a/(1+a)} > 0$ so that, based of Eq. (5), the sign for parameter $b$ determines the type of curvature: the surface is convex if $b > 0$ and concave if $b < 0$.

The location of the deviation peak is determined solely by the parameter $a$. It is in the half-height, $x/H = 1/2$, when $a = -2$, at $1/2 < x/H \le 1$ when $-2 < a < 0^-$, and at $0 \le x/H < 1/2$ when $-\infty < a < -2$. **Figure 3(b)** is an illustration of the deviation peak positions at different $a$. The ratio of maximum deviation to heap height has been normalized to unity.

Equation (5) states that the maximum deviation is determined by the parameters $a$ and $b$, while these two parameters depend on the angle of repose and the coefficient of friction which fulfills the condition $\mu < \tan \alpha$. For $\phi = 0.36$ (RCP) we get $2/(3(1-\phi)) = 1.04$ and for $\phi = 0.4$ (RLP) we get $2/(3(1-\phi)) = 1.11$. If we approximate $2/(3(1-\phi)) \approx 1$ for these two kinds of packing, we can write

$$\frac{z_{max}}{H} \approx \frac{1}{\tan \alpha} \left(\frac{\mu}{\tan \alpha - \mu}\right)^{\frac{\mu}{\tan \alpha - 2\mu}}. \tag{7}$$

**Figure 3(c)** and **(d)** is a plot of $z_{max}/H$ at various values of $\mu$ and $\tan \alpha$ ((c): surface plot and (d): contour plot) in the area that satisfies $\tan \alpha > \mu$. The minimum value of $z_{max}/H$ occurs around $\mu \approx \tan \alpha$ where a straight slope ($b = 0$) is formed (lower boundaries on two figures).

If we look at Eq. (7), it is as if $z_{max}/H \to \infty$ if $\mu \to (1/2) \tan \alpha$. This is not the case since Eq. (7) has been derived after approximating $2/(3(1-\phi)) \approx 1$ which is invalid for $\mu = (1/2) \tan \alpha$. For $\mu = (1/2) \tan \alpha$, based on Eqs. (3) and (4), we get $a = -\gamma/\rho g A = -2/3(1-\phi)$ and $b = (2 + \tan^2 \alpha)/\tan \alpha - 2/(3(1-\phi)\sin \alpha)$. Based on these two expressions, we get $z_{max}/H$ is finite. For example, if $\phi = 0.36$ and $\alpha = 40^o$, we obtain $a = -1.04167$ and $b = 1.8226$ so $z_{max}/H = 0.657$.

It can ve seen from **Fig. 3(c)**, $z_{max}/H$ is very large if $\mu \to 0$. In this condition, $a \to 0^-$, $(-a)^{-a/(1+a)} \to 1$, and $b \to \cot \alpha$ so that $z_{max}/H \to \cot \alpha$. Based on **Fig. 3(a)**, if $\mu \to 0$, $\alpha \to 0$ which causes $z_{max}/H \to \infty$. This means, the peak height is $H \to 0$. So, the reason for $z_{max}/H$ is very large when $\mu \to 0$, is not because $z_{max}$ is very large, but because $H \to 0$.

From Eq. (6), if $b = 0$ (straight slope), the friction coefficient reads

$$\mu_f = \frac{1}{1+\frac{1}{\cos^2\alpha}\left(\frac{2}{3(1-\phi)}-1\right)} \tan \alpha \tag{8}$$

By remembering $a < 0$, Eq. (6) gives $\mu < \mu_f$ if $b > 0$ and gives $\mu > \mu_f$ if $b < 0$. Equation (8) is challenged for further analysis. Since $\mu_f$ must be less than $\tan \alpha$, the second term in the denominator of Eq. (8) cannot be negative, or $\phi \geq 1/3$ which results in a critical value $\phi_c = 1/3$. Only when $\phi > \phi_c$, $\mu_f < \tan \alpha$ is satisfied to produce concave surfaces in which $\mu_f < \mu < \tan \alpha$. However, if $\phi < \phi_c$, only a convex surface ($b > 0$) can be generated. Because the porosities for both RCP and RLP packings are greater than $\phi_c$, both types of packing produce the convex surfaces. The concave surfaces are produced when the porosity is less than the porosity of the RCP packing and this is very rare. **Figure 4(a)** is the regions of the formation of convex and concave surfaces at various porosities.

In general, $\alpha \leq 45^o$ [15] so that, from Eq. (8), $\mu$ increases with $\alpha$ in this angle range. If $\alpha$ is slightly larger (smaller) than the critical angle of repose, the right side of Eq. (8) becomes larger (smaller) than $\mu_f$ so that a convex (concave) surface is formed. These results are illustrated in **Figs. 4(b)** and **(c)**. The straight profile occurs if the angle of repose is exactly equal the critical angle of repose. Thus, concave, convex, or straight profiles can occur in the same granular material.

The OA line in **Fig. 4(b)** has a slope angle equal to the critical angle of repose, $\alpha_0$. The OT line makes the true angle of repose ($\alpha > \alpha_0$) which connects the point on the baseline to the heap top. The OT curve represents the convex surface of the granular material. At the baseline, the angle formed is slightly greater than $\alpha$, and clearly greater than $\alpha_0$ (the angle of the line OB to the horizontal). The slope angle at this position can be assumed to be approximately equal to $\alpha_0 + \delta$ as reported by Jaeger et al [33].

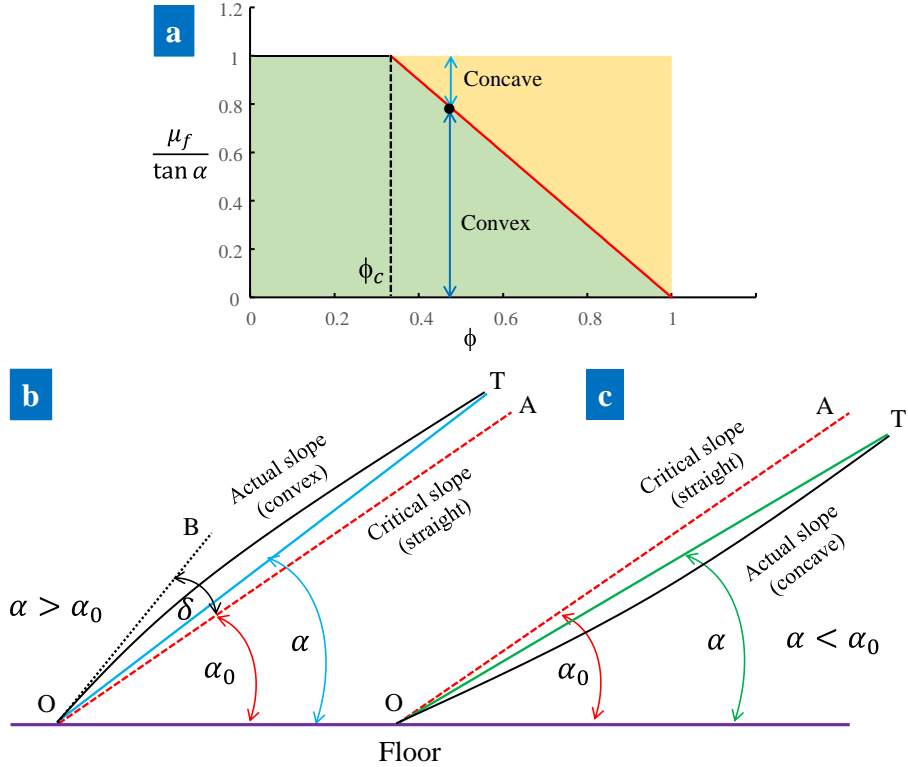

**Figure 4** (a) Locations where the surface is convex (green) and concave (yellow). The shape of the slope if the angle of repose is slightly different from the critical angle of repose. (b) The angle of repose is slightly larger than the critical angle of repose. (c) The angle of repose is slightly smaller than the critical angle of repose. See the explanation in the text.

The OA line in **Fig. 4(c)** has a slope angle equal to the critical angle of repose, $\alpha_0$. The OT line makes the true angle of repose ($\alpha < \alpha_0$) that connects the point on the baseline to heap top. The OT curve is the concave surface. At the baseline, the angle formed is slightly less than $\alpha$ and clearly smaller than $\alpha_0$. A slope angle that smaller than the critical angle of repose is permitted because it is stable.

We have strongly demonstrated, by modeling and experimentation, that the slopes of granular heaps and sand dunes surfaces are generally convex. Such convex profiles, indeed, have implicitly appeared in several previous reports as well as from photographs of sand dunes, but none has explicitly claimed them as the truly state. The results reported here are expected to open new understandings related to granular material properties as well as sand transport and geological records in several deserts in Africa, Asia, and America. Another fundamental result is that this model allows the estimation of the coefficient of friction between particles in the granular heap in the simplest way.

**Supplementary 1**

**Examples of Images of Sand Dunes**



Sergey Pesterev, Caravan in the desert, Morocco, Sahara (2013), Creative Commons License, https://commons.wikimedia.org/wiki/File:Caravan_in_the_desert.jpg, (Retrived on September 26, 2021)

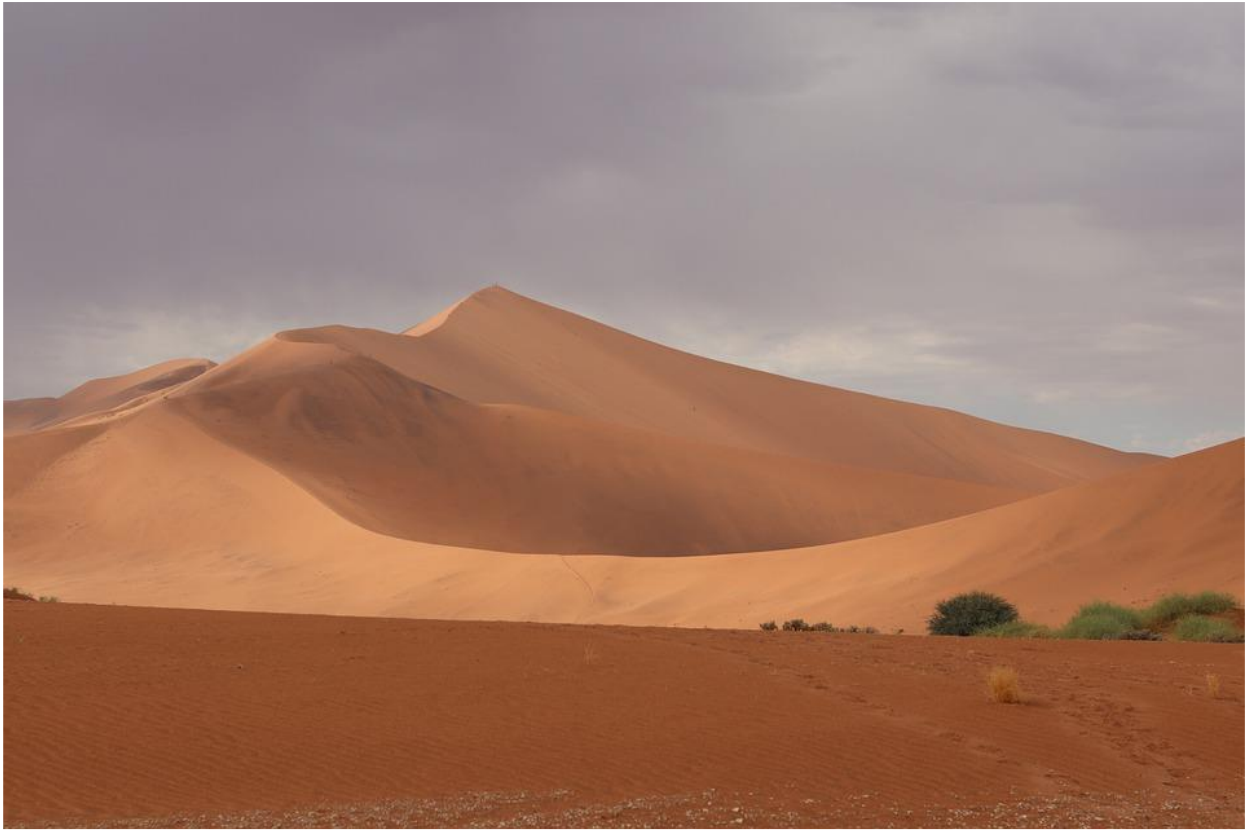

Dune Sand Namibia, image from From pixabay.com, Creative Commons License,

https://pixabay.com/photos/sossusvlei-namib-namibia-sand-dune-5339733/, (Retrived on September 26, 2021)

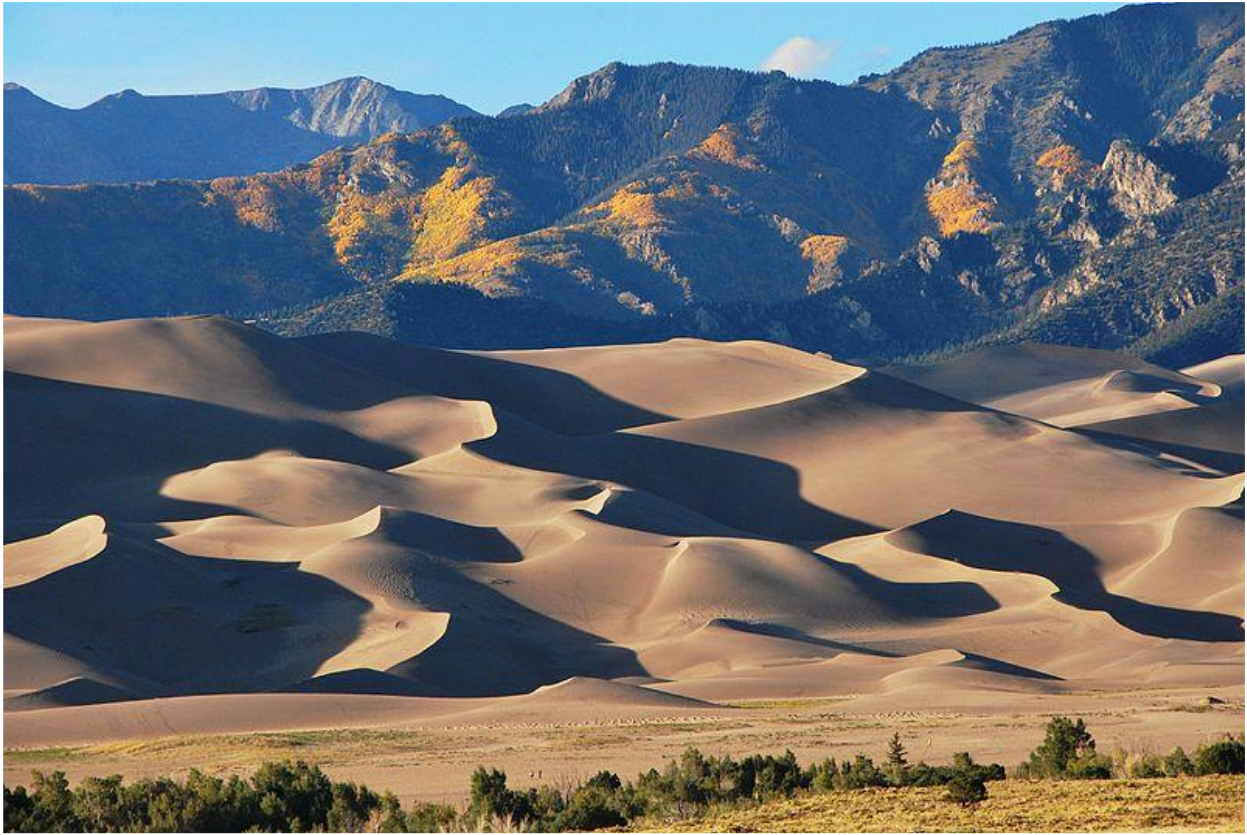

Great Sand Dunes National Park and Preserve (2014), Creative Commons License, https://commons.wikimedia.org/wiki/File:Gold_Aspens_Above_the_Dunes_at_Sunset_(15186713687).jpg (Retrived on September 26, 2021)

**Examples of Images of Granular Materials Heaps**

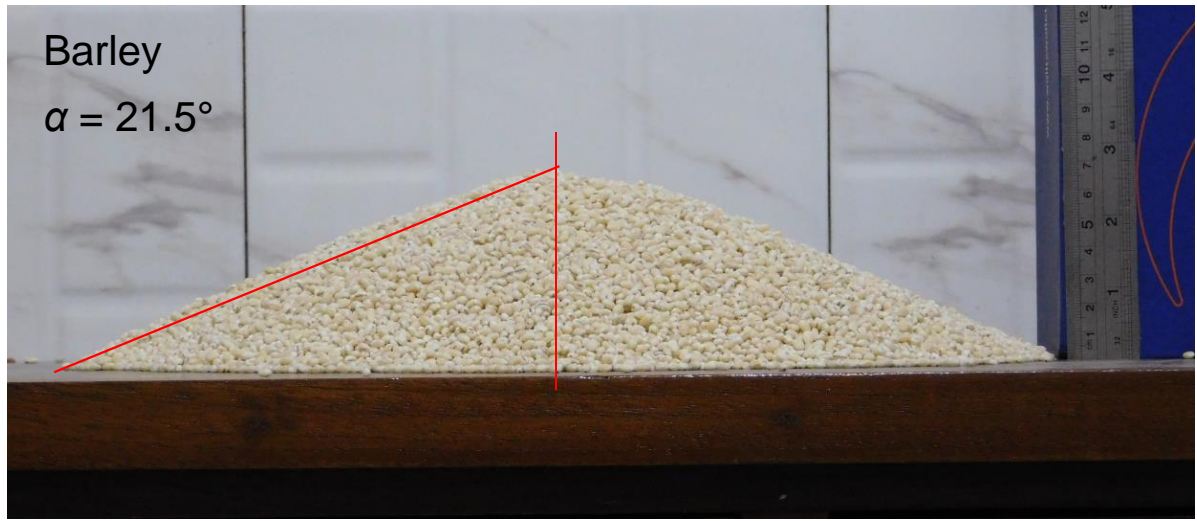

Barley
$\alpha = 21.5°$

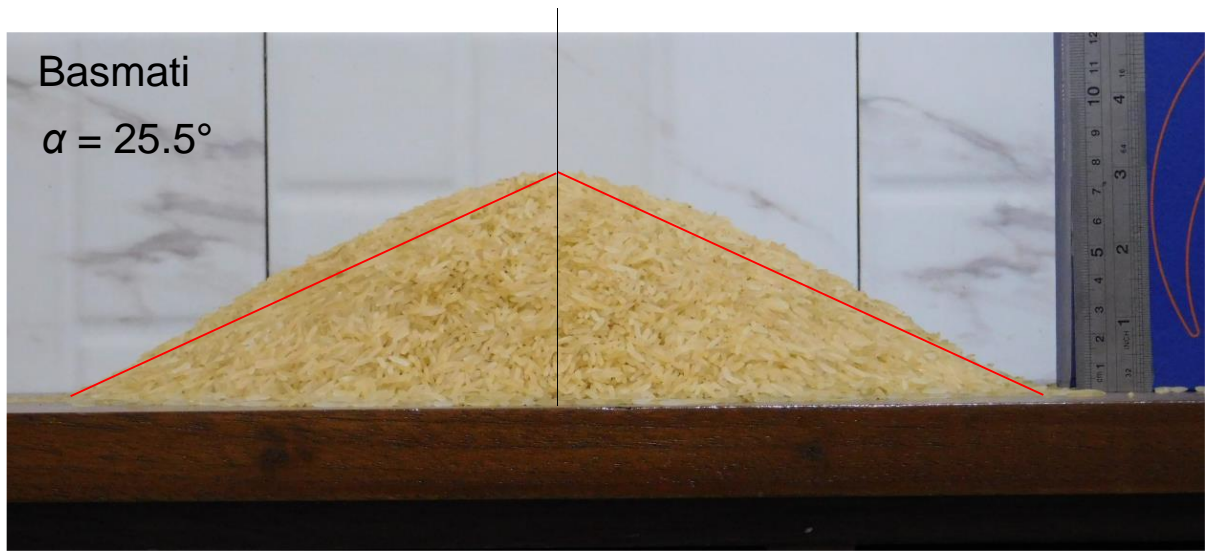

Basmati
$\alpha = 25.5°$

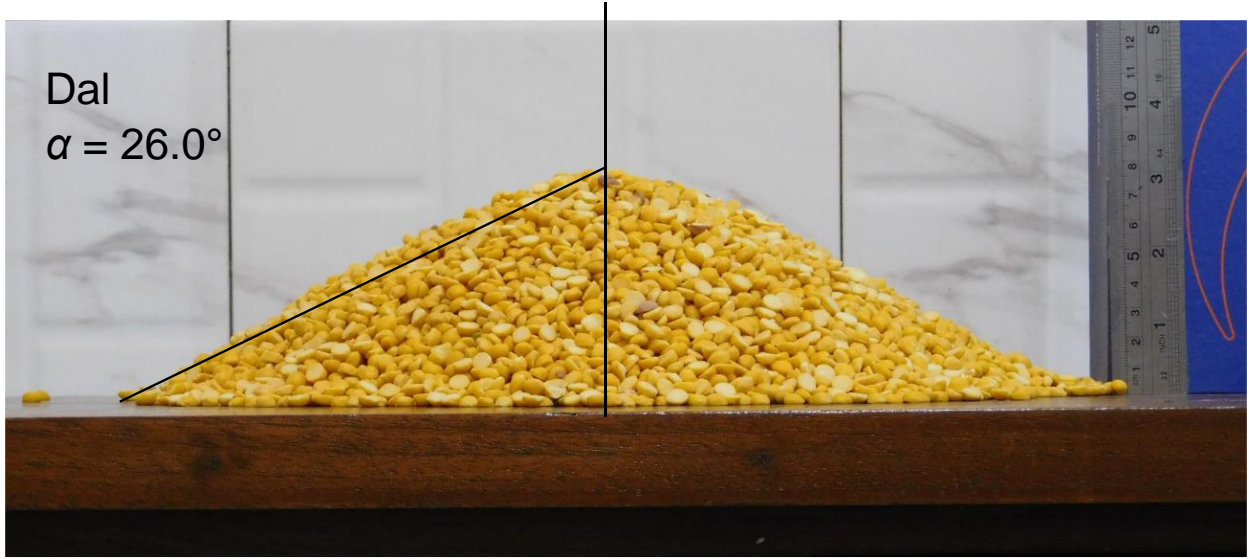

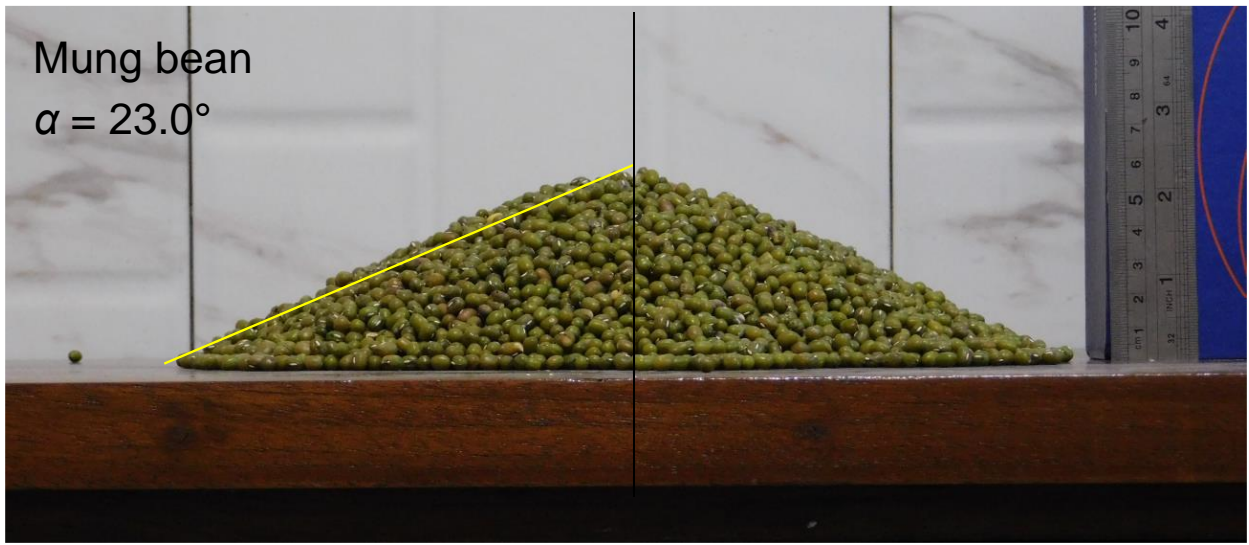

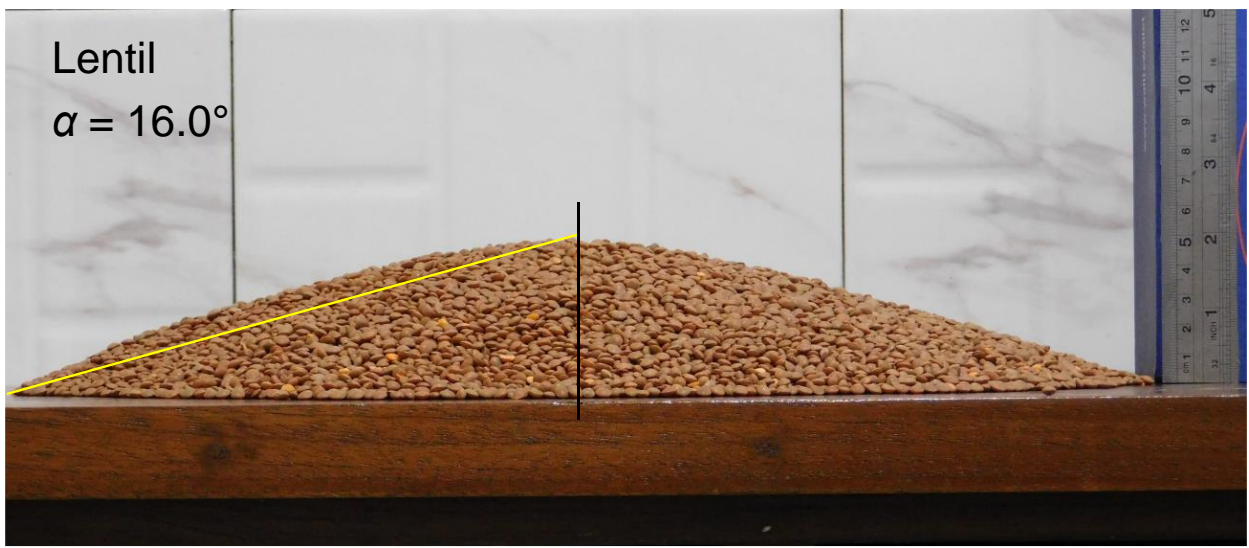

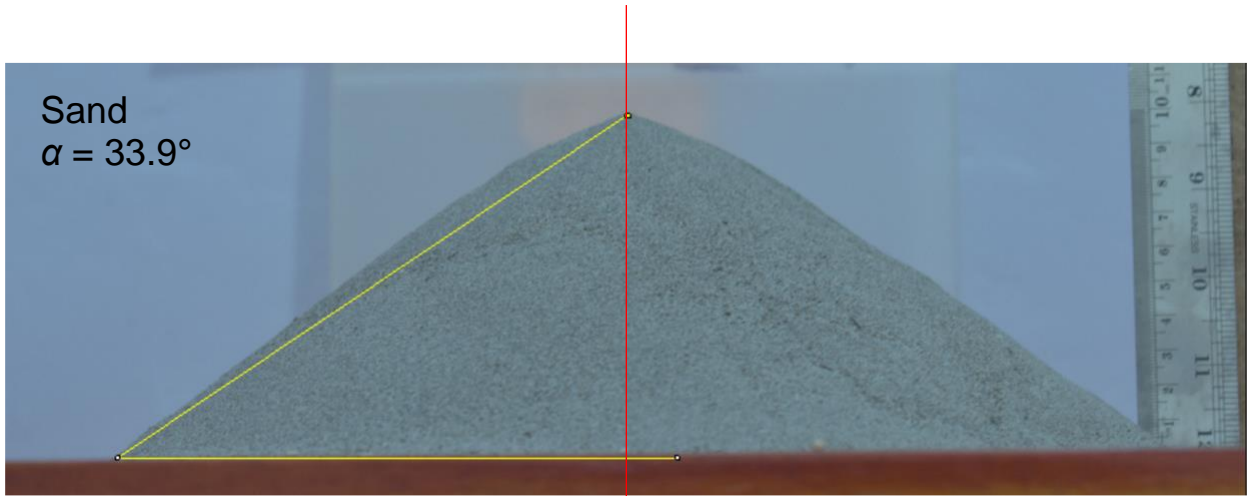

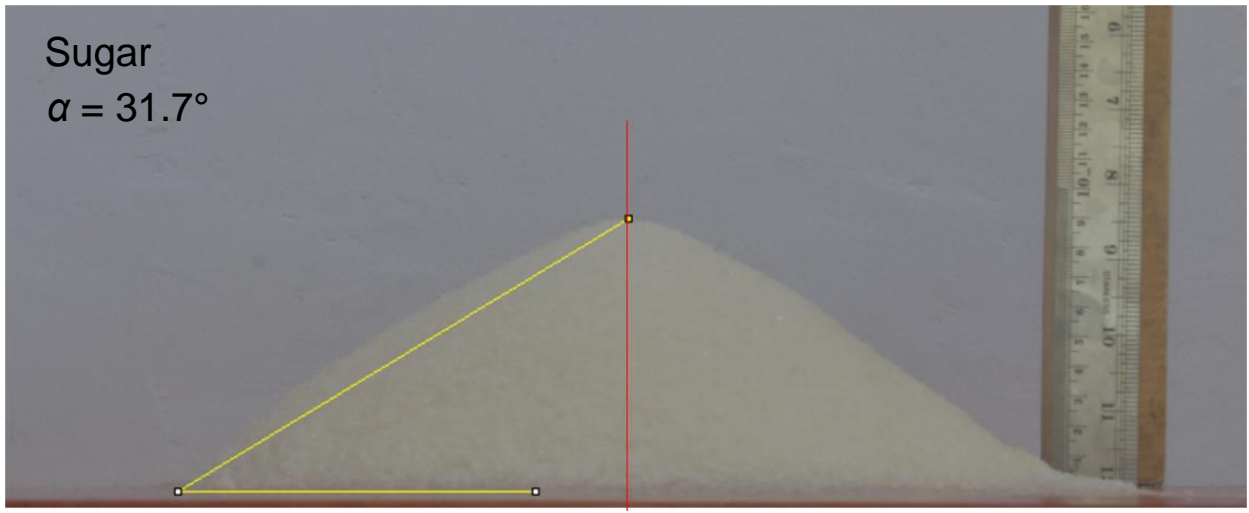

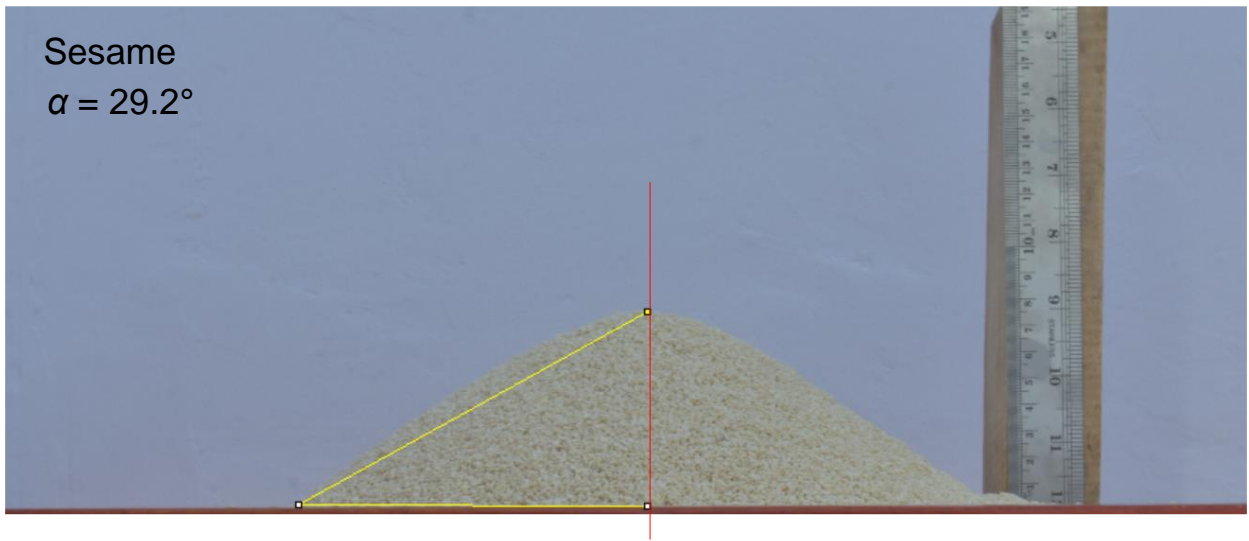

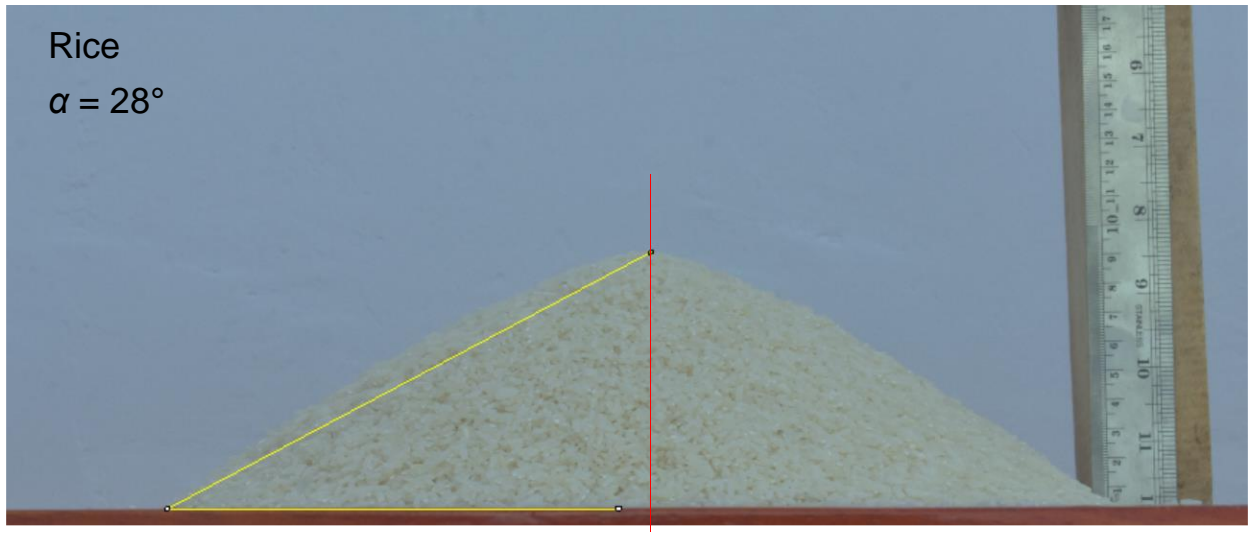

Rice
α = 28°

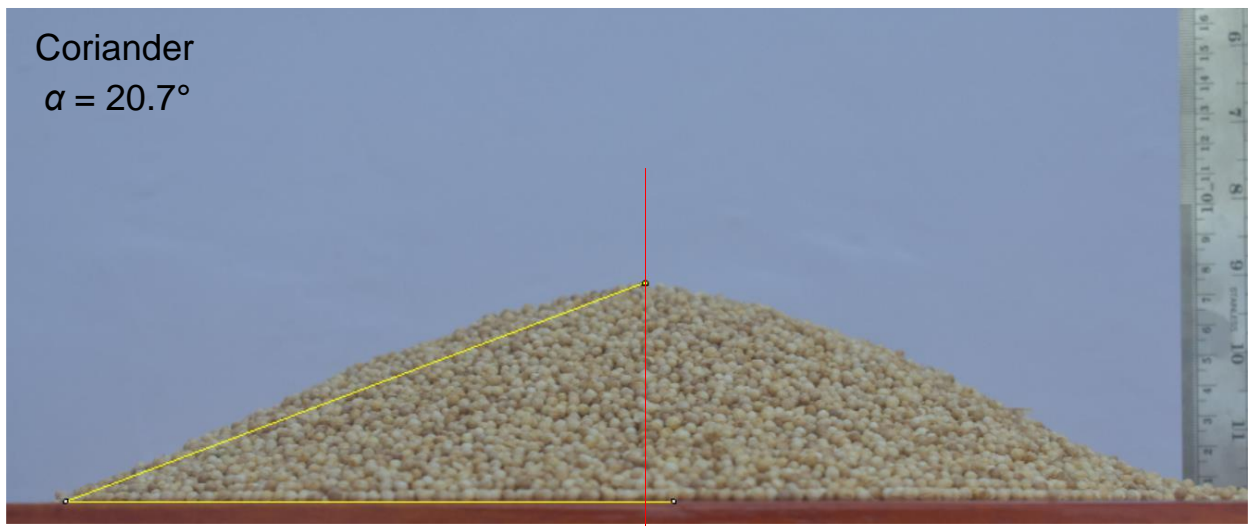

Coriander
α = 20.7°

**Supplementary 3**

**Derivation of Eq. (2)**

Consider one element on the granular surface of the material at a position between $x$ to $x + dx$ which contains only one particle. The horizontal direction forces equilibrium, $F(x + dx)\cos\theta(x + dx) + dN - F(x)\cos\theta(x) - dW = 0$, gives

$$dN = -\frac{d}{dx}[F(x)\cos\theta(x)]dx + dW \tag{S1}$$

The equilibrium of forces in the vertical direction, $F(x)\sin\theta(x) - F(x+dx)\sin\theta(x+dx) - df = 0$, gives

$$df = -\frac{d}{dx}[F(x)\sin\theta(x)]dx \tag{S2}$$

The frictional force $df$ originates from stress on the various contact sides of the particle and we assume that it is directed only inward (perpendicular to gravity As illustrated in the inset of **Figure 1**, the magnitude of the friction force $|df| = |\sum \sigma_i A_i|$ where $\sigma_i$ is the stress on the i-th contact surface and $A_i$ is the i-th contact surface area. If it is assumed that when the frictional force reaches its maximum value, the stress is directly proportional to the normal force, $\sigma_i = \kappa_i dN$, we can write $|df| = |\sum \kappa_i A_i dN| = \mu|dN|$, where $\mu = \sum \kappa_i A_i$ is the coefficient of frictionReferring to the Coulomb equation [S1], this decision was taken with the assumption that the cohesive forces were low enough to be neglected [S2]. From Eqs. (S1) and (S2) we get

$$\left|-\frac{d}{dx}[F(x)\sin\theta(x)]dx\right| = \left|-\mu\frac{d}{dx}[F(x)\cos\theta(x)]dx + \mu dW\right| \tag{S3}$$

The friction coefficient is generally contributed by the sliding and rolling of the particles. In the above equation we have assumed that only sliding motion contributes significantly. Rolling torque that contributes to rolling friction $\propto R$ (particle radius) [S3] so that for very small particles, its contribution can be neglected and only slip friction is considered to be contributing. To simulate the rolling effect, Zhou et al [S3] performed simulations on 10 mm granular particles, a size large enough that rolling cannot be neglected.

The force $F(x)$ acting along the surface arc can be approximated by a hydrostatic-like force [S4,S5] and all other forces have been accommodated in the frictional force with an inward direction perpendicular to gravity. The assumption of hydrostatic-like forces in granular systems has been widely reported [S6]. The hydrostatic pressure of the material satisfies the equation $P(x) = \rho g(H - x)$ where $\rho$ is the density of the granular system after taking into account the presence of pores and $g$ is the gravity acceleration. The hydrostatic force at position $x$ is $F(x) = \rho g A(H - x)$ where $A$ is the effective surface area. We approximate $A = \pi R^2$ where $R$ is the particle radius.

We don't yet know whether $\theta$ is monotonically increasing or monotonically descending, or oscillating when $x$ changes from 0 to $H$. However, from all observations that the slope of the granular material is almost constant (almost straight), the variation of $\theta$ to the value of $\bar{\theta}$ is not too large which causes $\sin\theta$ and $\cos\theta$ to be almost constant. Furthermore, $-\pi/2 < \theta < 0$ so that $\cos\theta > 0$ and $\sin\theta < 0$ which results in $F\cos\theta$ decreasing and $F\sin\theta$ increasing as $x$ gets bigger. Thus, the following two inequalities are satisfied $d/dx(F\sin\theta) > 0$ and $d/dx(F\cos\theta) < 0$. Since $dW/dx > 0$, Eq. (S3) can be written as $d/dx(F\sin\theta) = \mu dW/dx + \mu d/dx(F\cos\theta)$. By looking at **Fig. 1** on the right, we can write $dW = \gamma ds$ so that we get the following equation

$$\frac{d}{dx}F(\mu\cos\theta - \sin\theta) = -\mu\gamma\frac{ds}{dx} \tag{S4}$$

We can write the granular surface equation as $\psi(x) = y(x) + z(x)$ where $y(x) = (H - x)\tan\bar{\theta}$ which is the equation of the straight surface (mean) and $z(x)$ is the deviation of the straight surface, $|z(x)| \ll |y(x)|$ and $|dz/dx| \ll |dy/dx|$. We can write $\Delta s = \sqrt{\Delta x^2 + (\Delta y + \Delta z)^2}$ so we can easily prove that $ds/dx \approx 1/\cos\bar{\theta} + \sin\bar{\theta}\, dz/dx$. Equation (S4) can then be written

$$\frac{d}{dx}[F(\mu\cos\theta - \sin\theta) + \mu\gamma\sin\bar{\theta}\, z] = -\frac{\mu\gamma}{\cos\bar{\theta}} \tag{S5}$$

The general solution of Eq (S5) can be written as

$$F(\mu\cos\theta - \sin\theta) + \mu\gamma\sin\bar{\theta}\, z = \mu\gamma(H - x)/\cos\bar{\theta} + C \tag{S6}$$

where $C$ is a constant.

Next, let's write the angle as follows $\theta(x) = \bar{\theta} + \varphi(x)$, where $|\varphi| \ll 1$ and $|d\varphi/dx| \ll 1$. We use the trigonometric equality $\sin\theta = \sin(\bar{\theta} + \varphi) \approx \sin\bar{\theta} + \varphi\cos\bar{\theta}$, $\cos\theta = \cos(\bar{\theta} + \varphi) \approx \cos\bar{\theta} - \varphi\sin\bar{\theta}$ and approximation $\varphi \approx \tan\varphi = dz/dx$. Using these approximations, we can write $\sin\theta \approx \sin\bar{\theta} + \cos\bar{\theta}\, dz/dx$, $\cos\theta \approx \cos\bar{\theta} - \sin\bar{\theta}\, dz/dx$, and Eq. (S6) becomes

$$-(H - x)\frac{dz}{dx} + \frac{\mu\gamma\sin\bar{\theta}}{\omega(\mu\sin\bar{\theta} + \cos\bar{\theta})}z + \left(\frac{\mu\cos\bar{\theta} - \sin\bar{\theta}}{\mu\sin\bar{\theta} + \cos\bar{\theta}} - \frac{\mu\gamma}{\omega\cos\bar{\theta}(\mu\sin\bar{\theta} + \cos\bar{\theta})}\right)(H - x)$$

$$-\frac{C}{\rho g A(\mu\sin\bar{\theta} + \cos\bar{\theta})} = 0 \tag{S7}$$

For convenience, let's define $t = H - x$ so that $dz/dx = -dx/dt$ and Eq. (S7) simplifies to $tdz/dt + az + bt - c = 0$ where

$$a = \frac{\mu\gamma \sin\bar\theta}{\rho g A(\mu\sin\bar\theta + \cos\bar\theta)} = -\frac{\mu\gamma \cos\alpha}{\rho g A(-\mu\cos\alpha + \sin\alpha)} \tag{S8}$$

$$b = \frac{\mu\cos\bar\theta - \sin\bar\theta}{\mu\sin\bar\theta + \cos\bar\theta} - \frac{\mu\gamma}{\rho g A \cos\bar\theta(\mu\sin\bar\theta + \cos\bar\theta)} = \frac{\mu\sin\alpha + \cos\alpha}{-\mu\cos\alpha + \sin\alpha} - \frac{\mu\gamma}{\rho g A \sin\alpha(-\mu\cos\alpha + \sin\alpha)} \tag{S9}$$

$$c = \frac{C\prime}{\rho g A(-\mu\cos\alpha + \sin\alpha)} \tag{S10}$$

The general general solution for $z(t)$ is $z(t) = -bt/(1+a) + k_1 t^{-a} + c/a$ where $k_1$ is a constant. In order for the solution to be convergent in the range $0 \leq t \leq H$, $a$ must be negative. We also apply the boundary conditions $z(0) = z(H) = 0$ so that $c = 0$ which implies $C' = 0$ and $k_1 = bH^{1+a}/(1+a)$, and finally the compact form of the solution is given by

$$\frac{z(x)}{H} = \frac{b}{1+a}\left(\left(1 - \frac{x}{H}\right)^{-a} - \left(1 - \frac{x}{H}\right)\right). \tag{S11}$$

The convergence condition, $a < 0$, requires $-\mu\cos\alpha + \sin\alpha > 0$ (Eq. (S8)) which implies

$$\mu < \tan\alpha. \tag{S12}$$

Since $a < 0$, from Eq. (S11) we get $z(0) = z(H) = 0$.

[S5] R. Albert, M.A. Pfeifer, A.L. Barabási, and P. Schiffer, Slow drag in a granular medium. Phys. Rev. Lett. 82, 205–208 (1999).

# Supplementary 4

## Fitting of each curve in Fig 2.

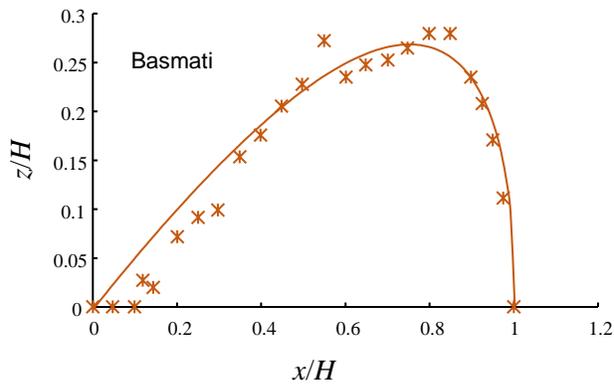
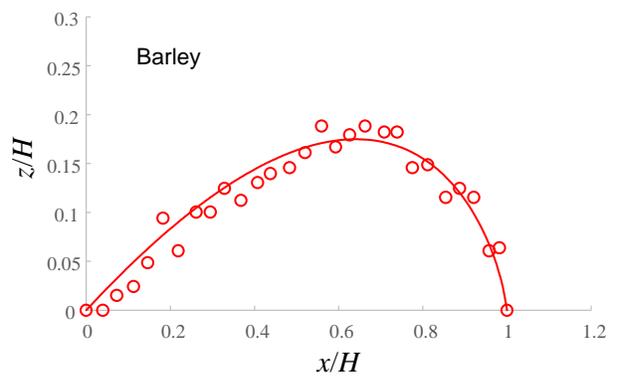
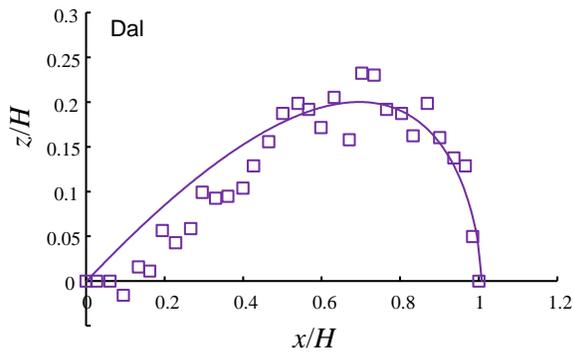
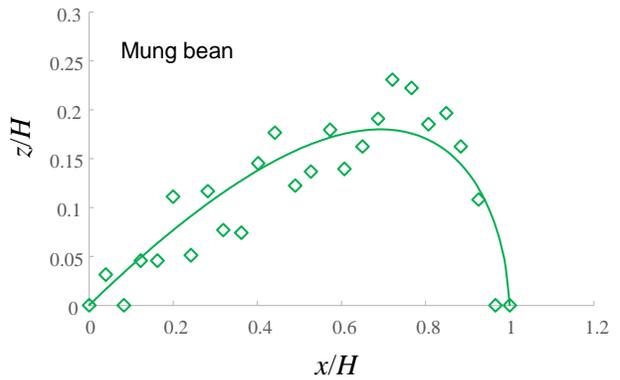

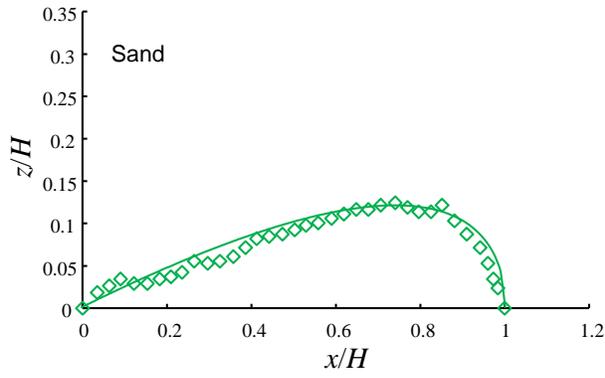
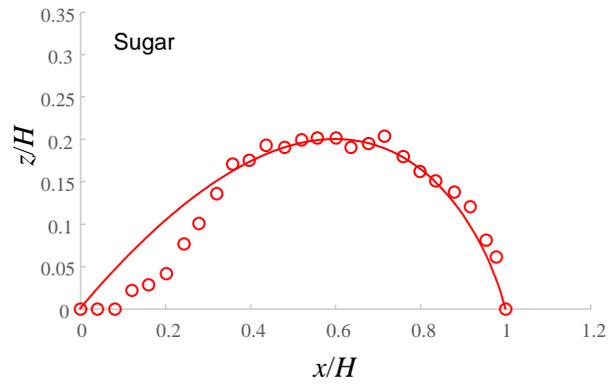
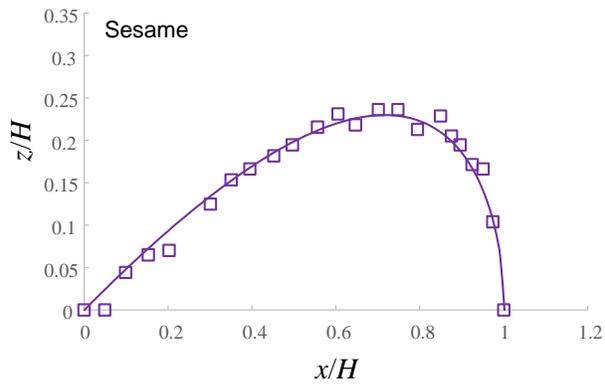
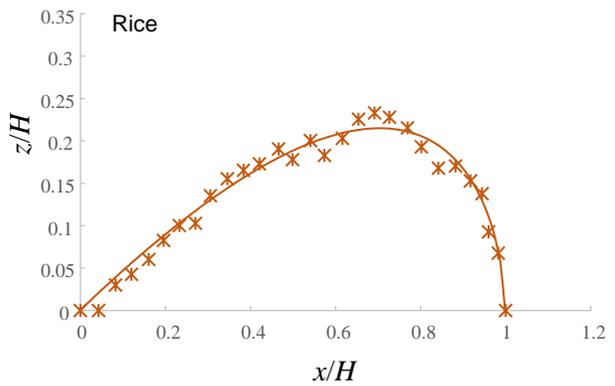
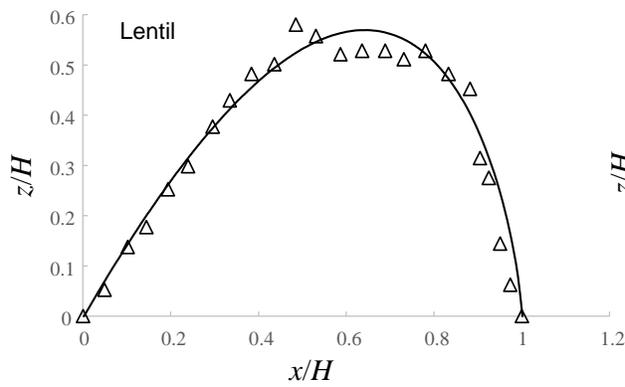
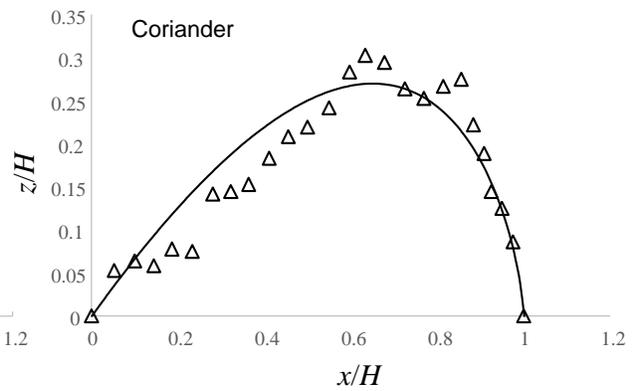

**Supplementary 5**

**Cummulative Distribution Function for Particles**

The cummulative distribustion function (CDF) for a lognormal distribution is

$$CFD(D) = \frac{1}{2}\left[1 + \text{Erf}\left(\frac{\ln D - \mu}{\sigma\sqrt{2}}\right)\right]$$

with $D$ is the particle diameter, $\mu$ and $\sigma$ are distribution parameters, and Erf(x) is the error function. We calculated $\mu$ and $\sigma$ using the following equations [https://en.wikipedia.org/wiki/Log-normal_distribution, retieved on September 26, 2021]

$$\mu = \ln\left(\frac{\bar{D}^2}{\sqrt{\bar{D}^2 + \text{Var}(D)}}\right)$$

$$\sigma^2 = \ln\left(1 + \frac{\text{Var}(D)}{\bar{D}^2}\right)$$

$$\bar{D} = \frac{1}{N}\sum_{i=1}^{N} D_i$$

$$\text{Var}(D) = \frac{1}{N}\sum_{i=1}^{N}(D_i - \bar{D})^2$$

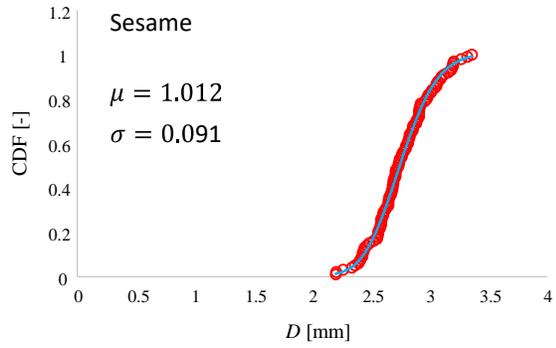
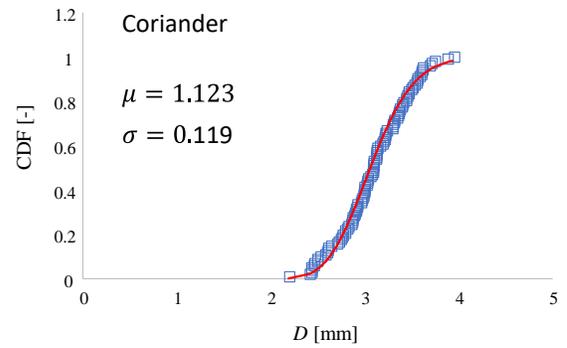
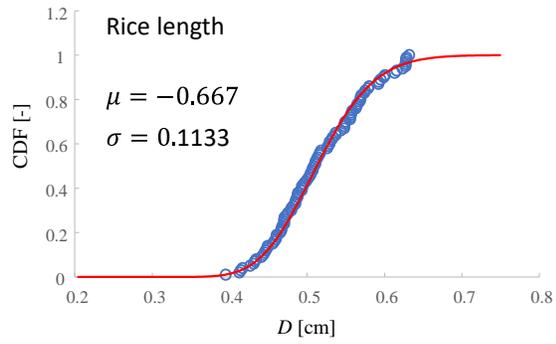
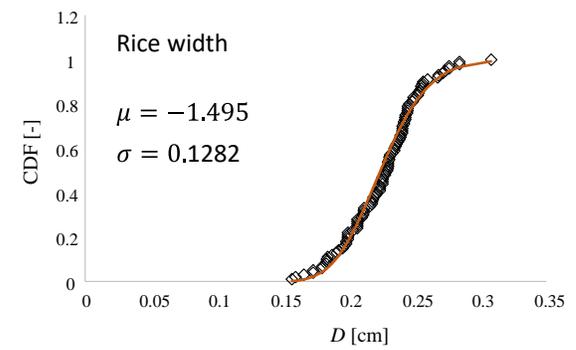
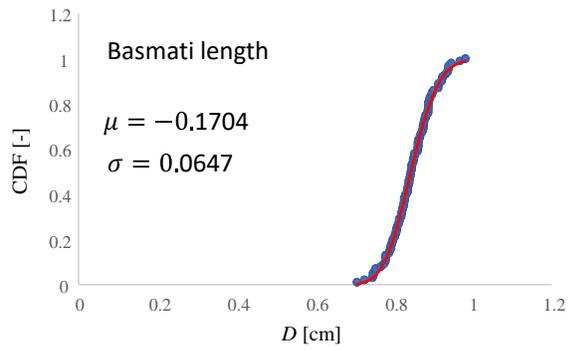
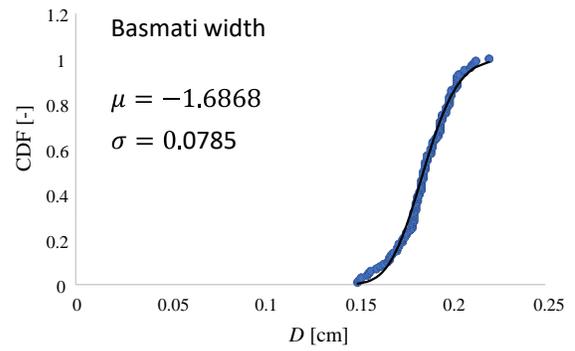
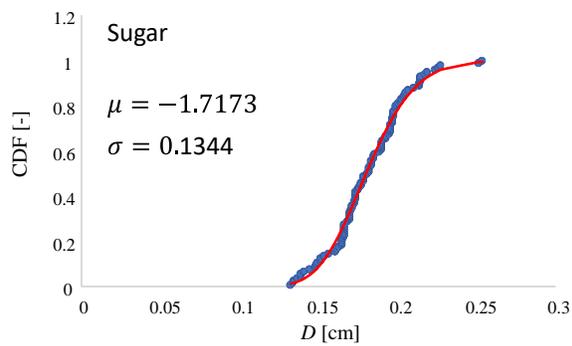
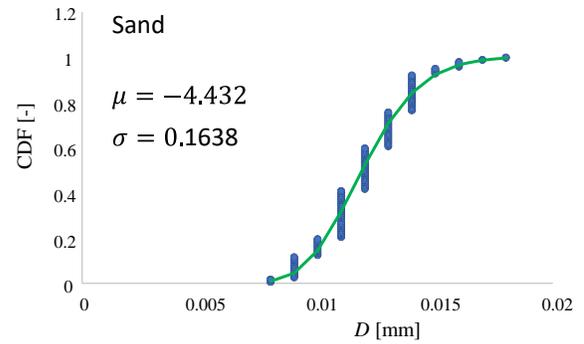

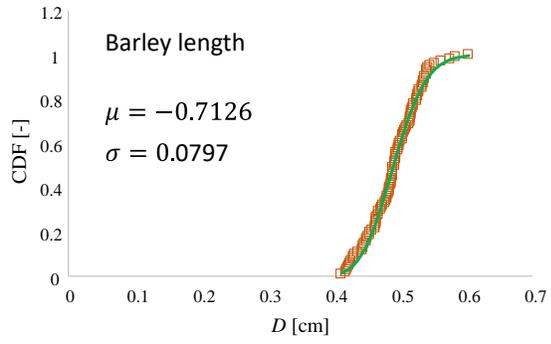
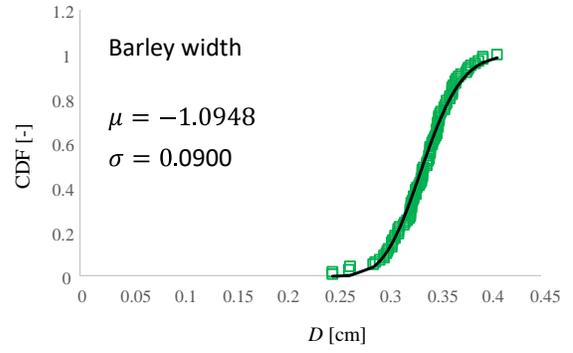
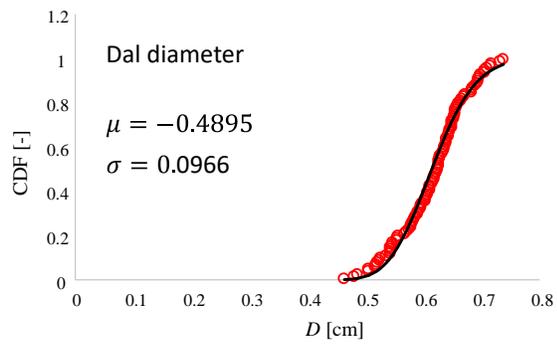
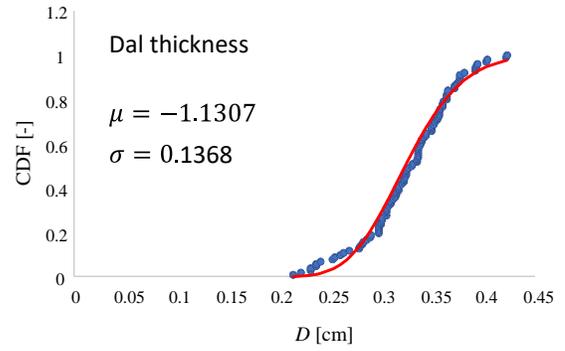
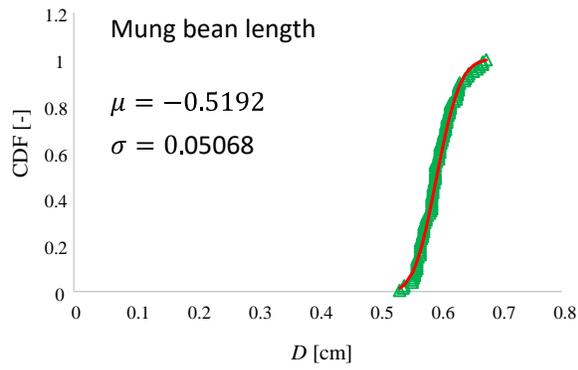
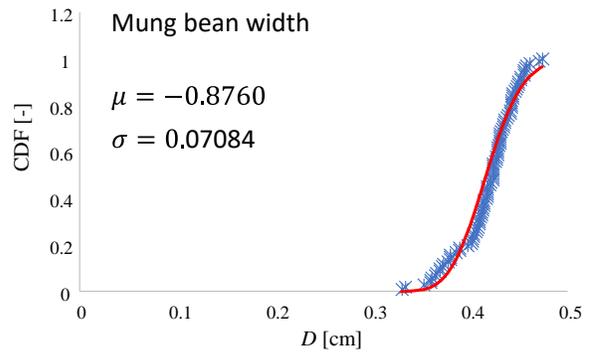
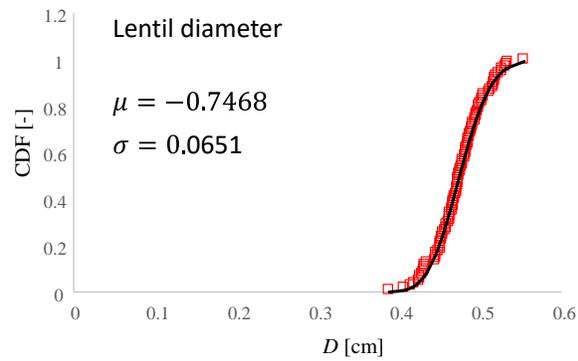
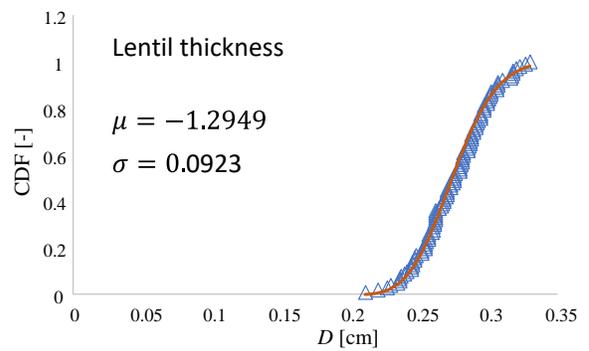